\documentclass[ aip, 
amsmath, amssymb,
preprint,
author-year,author-numerical]{revtex4-1}

\usepackage{graphicx}
\usepackage{float}
\restylefloat{figure}
\usepackage{subfig}

\usepackage{dcolumn}
\usepackage{bm}

\begin{document}


\title[Plane layer magnetoconvection]{Onset of plane layer magnetoconvection at low Ekman number}

\author{K\'elig Aujogue}
 \email{aujoguek@uni.coventry.ac.uk}

\author{Alban Poth\'erat}%

\affiliation{ 
Coventry University 
}%

\author{Binod Sreenivasan}

\affiliation{%
Centre for Earth Sciences, Indian Institute of Science, Bangalore 560 012, India.
}%

\date{\today}


\begin{abstract}

 We study the onset of magnetoconvection between two infinite horizontal 
planes subject to a vertical magnetic field aligned with background rotation. 
In order to gain insight into the convection taking place in the Earth's tangent 
cylinder (TC), we target regimes of asymptotically strong rotation. The 
critical Rayleigh  number $Ra_{c}$ and critical wavenumber are computed 
numerically by solving the linear stability problem in a systematic way. 
A parametric study is conducted, varying the Ekman number, 
$E$ (ratio of viscous to Coriolis forces) and the Elsasser number, 
$\text{\ensuremath{\Lambda}}$ (ratio of the Lorentz 
force to the Coriolis force). $E$ is varied from $10^{-9}$
to $10^{-2}$ and $\Lambda$ from $10^{-3}$ to $1$. 
Apply to arbitrary thermal and magnetic Prandtl numbers, our results verify and confirm previous experimental and theoretical results showing 
the existence of two distinct unstable modes at low values of $E$ -- one being 
controlled by the magnetic field, the other being controlled by viscosity (often called the viscous mode).
Asymptotic scalings for the onset of these modes have been numerically confirmed and 
precisely quantified. We show that with no-slip boundary conditions, 
the asymptotic behaviour is reached for $E<10^{-6}$ 
and establish a map in the $(E,\Lambda)$ plane. We distinguish regions where 
convection sets in either in the magnetic mode or in
the viscous mode. Our analysis gives the regime in which
the transition between magnetic 
and viscous modes may be observed. We also show 
that within the asymptotic regime, the role played by the kinematic boundary 
conditions is minimal. 
\end{abstract}

\pacs{Valid PACS appear here}
\maketitle


\section{\label{sec:level1}Introduction}

In this paper, we analyse the onset of plane-layer convection governed by
the interplay between the magnetic (Lorentz), buoyancy and Coriolis forces, to obtain an insight
into how convection in the Tangent Cylinder (TC) region
of the Earth's liquid core is driven. This region is bounded by the Earth's
solid inner core at its bottom, the mantle at its top, and by an
imaginary cylinder tangent to the solid inner core and parallel to the Earth's
rotation axis. Intense convection, compositional and thermal, is believed to take
place in this region, affecting the structure of the
magnetic field near the poles [\onlinecite{sreenivasan2006azimuthal}].
The Earth's self-generated magnetic field is thought 
to affect the structure
of convective cells in the TC, producing strong anticyclonic polar vortices that show
up in the secular variation of the geomagnetic field  [\onlinecite{05grl}].
 The aim of our study is to
find out whether onset of convection is sensitive to the Lorentz force in the regime
of strong rotation that characterises the Earth.

Previous work on plane rotating magnetoconvection has been motivated
either by geophysical or engineering applications involving liquid
metals [\onlinecite{burr2001rayleigh,busse1982stability,cioni2000effect,houchens2002rayleigh,takashima1999buoyancy,volz1999thermoconvective,yanagisawa2013convection}].
A number of geophysically motivated studies focused on the dynamics
outside the TC: an early study [\onlinecite{fearn1979thermally}] derived
theoretical scalings for the critical Rayleigh number and wave number at
the onset of convection as a function of the magnetic field intensity
and magnitude of the Coriolis force. Other studies
[\onlinecite{busse1970thermal,busse1982stability,carrigan1983experimental}]
showed experimentally and theoretically that, in this region convection
and rotation generated tall columns parallel to the rotation axis. A recent study
investigated the role of a dipolar magnetic field in enhancing helicity in convection 
columns [\onlinecite{sreenivasan2011helicity}], which can explain subcritical behaviour
as well as the preference for the axial dipole in rapidly rotating dynamos. 
These studies, however, do not consider the particularity of the TC, which, though imaginary,
acts somewhat as a physical boundary because the presence of the solid inner core
makes overcoming the Taylor-Proudman constraint more difficult. When convection
does set in, motions vary strongly along $z$ as heat and composition flux
have a substantial component in the $z$-direction.
Due to the large aspect ratio of the TC, the curvature of the
top and the bottom boundaries are not expected to play a lead role, at
least at the onset of convection. On these grounds, a simple plane
geometry is expected to provide a fair, albeit simplistic, representation
of the TC. In this geometry, it was theorised  [\onlinecite{chandrasekhar1961hydrodynamic},
\onlinecite{podvigina2010stability}] and experimentally observed
[\onlinecite{nakagawa1957experiments}] that the convection could set off
through an instability either of a magnetic or a viscous mode, depending
on the values of the Ekman number (Viscous
to Coriolis forces) and of the Elsasser number (Lorentz to Coriolis
forces). While the magnetic mode has a low horizontal wavenumber,
the viscous mode is characterised by thin structures of high horizontal
wavenumber parallel to the rotation axis. One would expect that convective flows driven by these
two mechanisms to differ significantly. These studies showed that
transition between these modes resulted in a brutal change in the
wavelength of the observed convective pattern, but concerned only
large values of $E$ ($>10^{-5}$). Such values may be too far from the
asymptotic regimes relevant to the Earth's ($E\sim10^{-14}$) [\onlinecite{gubbins2001rayleigh}]
to be applicable to it.

There have been experimental studies dedicated to the dynamics of the TC for $E=10^{-3}$
[\onlinecite{aurnou2001experiments,aurnou2003experiments}], but in the absence of
the magnetic field, only the viscous mode of convection could be observed. 
The link between plane layer magnetoconvection and convection in the Earth's TC was
first established by linear onset calculations as well as numerical
simulations of the geodynamo [\onlinecite{05grl,sreenivasan2006azimuthal}], where
substantial thickening of buoyant plumes under the effect of
the magnetic field was noted, albeit at values of $E$ down to $10^{-4}$ only.
Crucially, these studies showed that 
non-axisymmetric, Earth-like polar vortices are obtained only
through the action of the magnetic field.  

To explore Earth-relevant regimes, we look at plane-layer magnetoconvection
at values of $E$ low enough to
find an asymptotic regime. Although actual regimes of the TC remain
beyond the reach of this analysis, asymptotic scalings are relevant
to it. In the same spirit, we shall characterise the consequence of
using either a no-slip boundary condition or its less computationally
demanding stress-free counterpart on these regimes.

The paper is organised as follows: Section 2 introduces the governing
equations and the numerical method to solve them is validated. Results
are discussed in Section 3 results and scalings for the asymptotic
regimes are expressed in terms of $E$ and $\Lambda$. Relevance to
the Earth is discussed in section 4.


\section{Governing Equations and Numerical Method}

\subsection{Governing equations}

\begin{figure}
\includegraphics{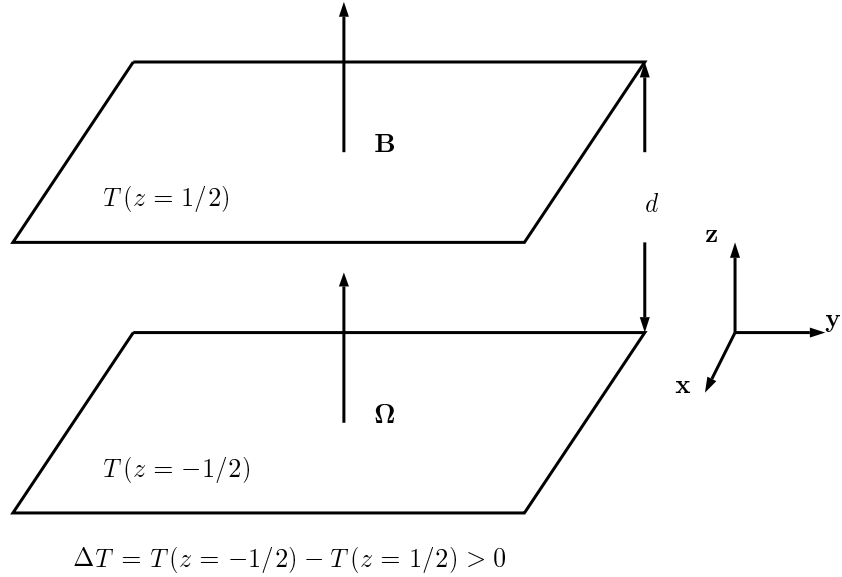}
\caption{\label{fig:Schematic-illustration-of}Schematic illustration of the
geometry}
\end{figure}

We consider an incompressible fluid (viscosity $\nu$, thermal diffusivity
$\kappa$, magnetic diffusivity $\eta$, density $\rho$, expansion
coefficient $\alpha$) confined between two differentially heated infinite
horizontal plane boundaries, separated by a distance $d$. The temperature
difference between them is $\Delta T$. The flow rotates at a speed
$\Omega$ around the vertical axis $\mathbf{z}$ and is permeated by 
a uniform vertical magnetic field $\mathbf{B}=B_{0}\mathbf{e_{z}}$.
Figure \ref{fig:Schematic-illustration-of} illustrates our geometry.

The flow is governed by the full incompressible MHD equations under
the Boussinesq approximation. Normalising
lengths by $d$, the velocity by $\eta/d$, the pressure by
$\rho\eta\Omega$, the magnetic field by $B_{0}$, the time by $d^{2}/\eta$,
the temperature by $\Delta T$ and the rotation speed by $\Omega$, the
equations can be written in non-dimensional form, as follows:

\begin{equation}
P\! m^{-1}E(\frac{\partial\mathbf{u}}{\partial t}+(\mathbf{u}\cdot\boldsymbol{\nabla})\mathbf{u})+2\mathbf{\hat{z}}\times\mathbf{u}=-\boldsymbol{\nabla}p+R\! aT+\Lambda(\mathbf{\boldsymbol{\nabla}}\times\mathbf{B})\times\mathbf{B}+E\nabla^{2}\mathbf{u},\label{eq:}
\end{equation}

\begin{equation}
\frac{\partial\mathbf{B}}{\partial t}=\boldsymbol{\nabla}\times(\mathbf{u}\times\mathbf{B})+\nabla^{2}\mathbf{B},\label{eq:-1}
\end{equation}

\begin{equation}
\frac{\partial T}{\partial t}+(\mathbf{u}\cdot\boldsymbol{\nabla})T=P\! mP\! r^{-1}\nabla^{2}T,\label{eq:-2}
\end{equation}

\begin{equation}
\mathbf{\mathbf{\boldsymbol{\nabla}}}\cdot\mathbf{u}=0,\label{eq:-3}
\end{equation}

\begin{equation}
\boldsymbol{\nabla}\cdot\mathbf{B}=0,\label{eq:-4}
\end{equation}

where $\mathbf{B}$ is the total magnetic field. The system is controlled
by 5 non-dimensional parameters: the Ekman number, $E=\nu/\Omega d^{2}$, 
a modified Rayleigh number, $R\! a=g\alpha\Delta Td/\eta\Omega$,
the Elsasser number, $\Lambda=B^{2}/\mu_{0}\eta\rho\Omega$, the
Prandtl number, $P\! r=\nu/\kappa$ and the magnetic Prandtl
number, $P\! m=\nu/\eta$. We applied two different kinds of
boundary conditions: 
stress-free magnetic (SFM) (conditions (\ref{eq:condi1})-- 
(\ref{eq:condi4}) below) and no-slip magnetic (NSM) (conditions (\ref{eq:condi3})-- 
(\ref{eq:condi5}) below). These are given respectively for $z=\pm1/2$ as:
\begin{equation}
u_z=0\quad\mathrm{\text{(impermeability)},}\label{eq:condi1}
\end{equation}

\begin{equation}
\dfrac{d^2 u_z}{dz^2}=0\quad\mathrm{\text{(stress-free)}},\label{eq:condi2}
\end{equation}

\begin{equation}
(\boldsymbol{\nabla}\times\mathbf{B}).\mathbf{e_{z}}=0\quad\mathrm{\text{(electrically insulating)},}\label{eq:condi3}
\end{equation}

\begin{equation}
T(z=-1/2)=1,\; T(z=+1/2)=0,\label{eq:condi4}
\end{equation}

\begin{equation}
\mathbf{u}=0\quad\mathrm{\text{(no-slip)}.}\label{eq:condi5}
\end{equation}

For both sets of boundary conditions, the system has a simple solution
with $\mathbf{u}_{0}=0$, $\mathbf{B}_{0}=0$, and $T=T_{0}+z\Delta T$.
We are interested in the linear stability of this basic state. The
problem's invariance in the $x$ and $y$ directions allows us to decompose 
all physical quantities as 
$g(z)=g_{0}+\hat{g}(z)e^{i\mathbf{a}.\mathbf{r}_{\perp}}$,
where $\mathbf{r}_{\perp}=(x,y)$ and $\mathbf{a}$ is the wave number.
Following Sreenivasan \& Jones [\onlinecite{sreenivasan2006azimuthal}],
we shall only seek the shape of the unstable modes, not their growth
rate. The perturbation equations in steady state are given by

\begin{equation}
E(D^{2}-a^{2})\hat{\omega_{z}}+2D\hat{u_{z}}+\Lambda D\hat{j_{z}}=0,\label{eq:-7}
\end{equation}

\begin{equation}
E(D^{2}-a^{2})^{2}\hat{u_{z}}-2D\hat{\omega_{z}}+\Lambda(D^{2}-a^{2})D\hat{b_{z}}-2R\! a\hat{T'}=0,\label{eq:-8}
\end{equation}

\begin{equation}
(D^{2}-a^{2})\hat{b_{z}}+D\hat{u_{z}}=0,\label{eq:-9}
\end{equation}

\begin{equation}
(D^{2}-a^{2})\hat{j_{z}}+D\hat{\omega_{z}}=0,\label{eq:-10}
\end{equation}

\begin{equation}
P\! mP\! r^{-1}(D^{2}-a^{2})\hat{T'}+\hat{u_{z}}=0.\label{eq:-11}
\end{equation}

Here $D$ is the derivative along $\mathbf{z}$,  $\hat{\omega_{z}}$,
$\hat{u_{z}}$, $\hat{j_{z}}$ and  $\hat{b_{z}}$ are the $z$-components of the
vorticity, velocity, electric current and magnetic field perturbations and $\hat{T^{'}}$
is the temperature perturbation.
The nondimensional wave number is denoted by $a=\parallel\mathbf{a}\parallel$.
Eq. $(\ref{eq:-7})$ is obtained from $\boldsymbol{\nabla}\times(\ref{eq:}) \cdot \mathbf{e_{z}}$,
$(\ref{eq:-8})$ from $\boldsymbol{\nabla}\times [\boldsymbol{\nabla}\times(\ref{eq:})] \cdot
\mathbf{e_{z}}$,
$(\ref{eq:-9})$ from $\boldsymbol{\nabla}\times(\ref{eq:-1}).\mathbf{e_{z}}$,
$(\ref{eq:-10})$ as $\boldsymbol{\nabla}\times [\boldsymbol{\nabla}\times(\ref{eq:-1})] \cdot
\mathbf{e_{z}}$
and Eq. $(\ref{eq:-11})$ follows from Eq. $(\ref{eq:-2})$. The boundary conditions
$(\ref{eq:condi1})$--$(\ref{eq:condi5})$ take the form
\begin{equation}
D^{2}\hat{u_{z}}=\hat{u_{z}}=D\hat{\omega_{z}}=\hat{j_{z}}=\hat{T'}=0\quad \mathrm{for} \quad z=\pm1/2,
\quad (\mathrm{SFM}) \label{eq:-12}
\end{equation}
\begin{equation}
D\hat{u_{z}}=\hat{u_{z}}=\hat{\omega_{z}}=\hat{j_{z}}=\hat{T'}=0 \quad \mathrm{for} 
\quad z=\pm1/2\:. \quad (\mathrm{NSM}) \label{eq:-13}
\end{equation}
The problem becomes a generalized eigenvalue problem of the form $A X=Ra B X$. 
The critical Rayleigh number for the onset of convection $R\! a_{c}$
is found as an eigenvalue of the problem for any given $a$ and minimised
over $a$ as in [\onlinecite{chandrasekhar1961hydrodynamic}]. With help of  a formal transformation of $\hat{T'}$ as 
$\hat{T_m'}=P\! mP\! r^{-1}\hat{T'}$
and $R\! a_{c}$ as $R\! a_{cm}=P\! rP\! m^{-1}R\! a_{c}$, the solution of this model is made independent of the magnetic and thermal diffusivities. The results presented there after therefore extend to arbitrary values of $P\!m$ and $P\!r$.

\subsection{Numerical method}

\begin{figure}
\includegraphics[scale=0.4]{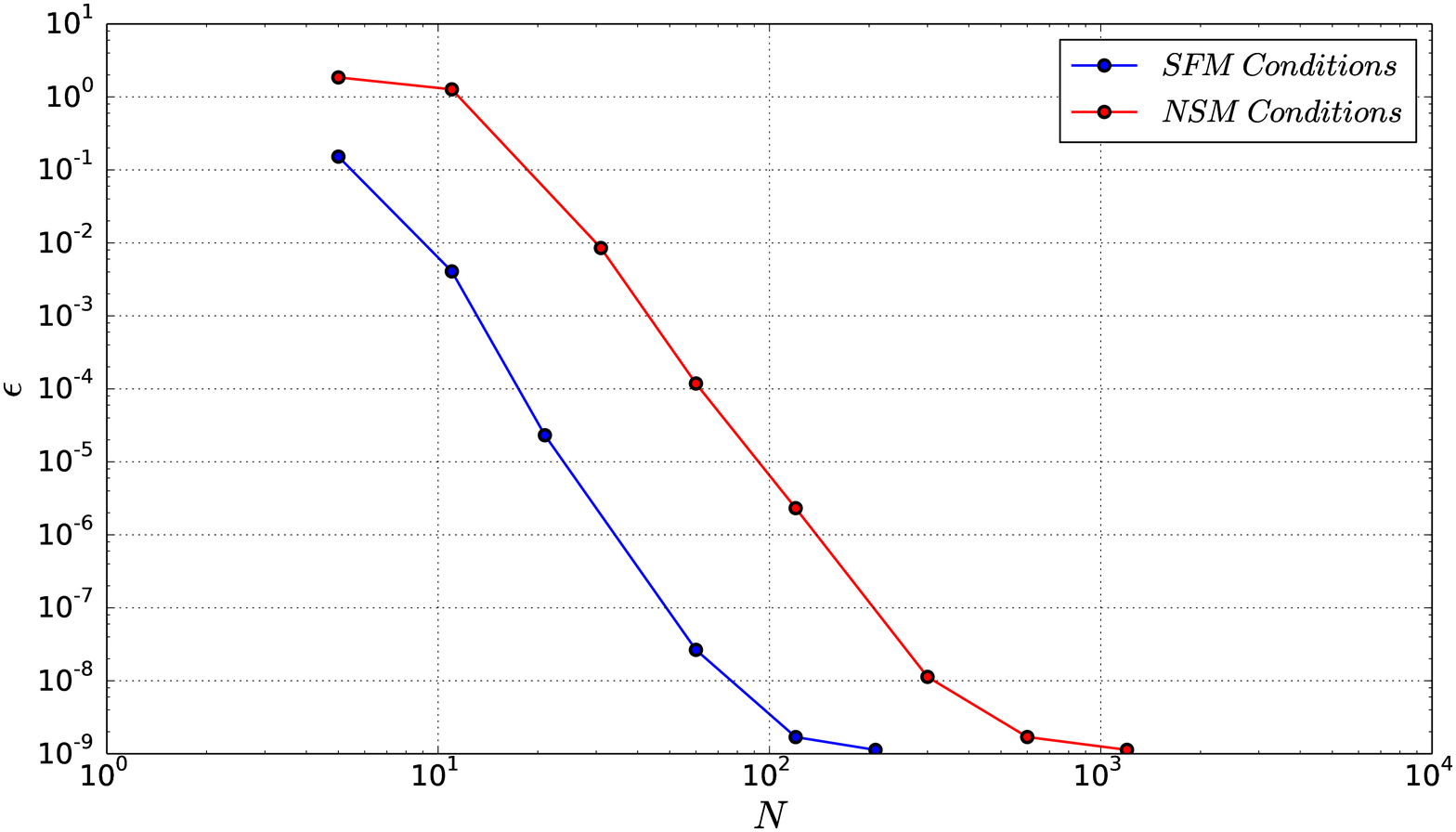}
\caption{\label{fig:Convergence-test}Convergence test}
\end{figure}

Eqs. $(\ref{eq:-7})-(\ref{eq:-11})$ were solved numerically
using a spectral collocation method based on Chebyshev polynomials
[\onlinecite{schmid2001stability}]. In the no-slip case, a boundary layer
of thickness $\delta=2\sqrt{E\pi}$ develops along the walls [\onlinecite{1973RPPh...36..159A}],
and we have ensured that at least 3 collocations points were in it.
Some convergence tests have been performed to ensure that the resolution
is adequate. The results are presented in figure \ref{fig:Convergence-test},
where we varied the number of collocations points $N$ between $5$
to $3000$. In the SFM case, the tests were performed for
$\Lambda=1,\quad E=10^{-9}\quad\mathrm{and}\quad a=3.149$. NSM conditions
were tested with $\Lambda=1,\quad E=10^{-7}\quad\mathrm{and}\quad a=3.333$.
We chose these parameters to ensure a good convergence at
the lowest $E$ we investigated. We look at the value of the error, $\epsilon$ on $R\! a_{c}$ 
relative to its value obtained for $N=3000$. For both types of boundary conditions,
$N>100$ gives a small relative error. On the basis of this
test, the results presented in the next section have been obtained with
$N=600$ for the SFM case and $N=1200$ for the NSM case. 

We performed a parametric study with $a=\left[1,1500\right]$, $E=\left[10^{-9},10^{-2}\right]$
and $\Lambda=\left[10^{-3},2\right]$ for SFM and $a=\left[1,1500\right]$,
$E=\left[10^{-8},10^{-2}\right]$ and $\Lambda=\left[10^{-3},2\right]$ for 
NSM.


\section{Results}

\subsection{General properties}

\begin{figure}
\includegraphics[scale=0.45]{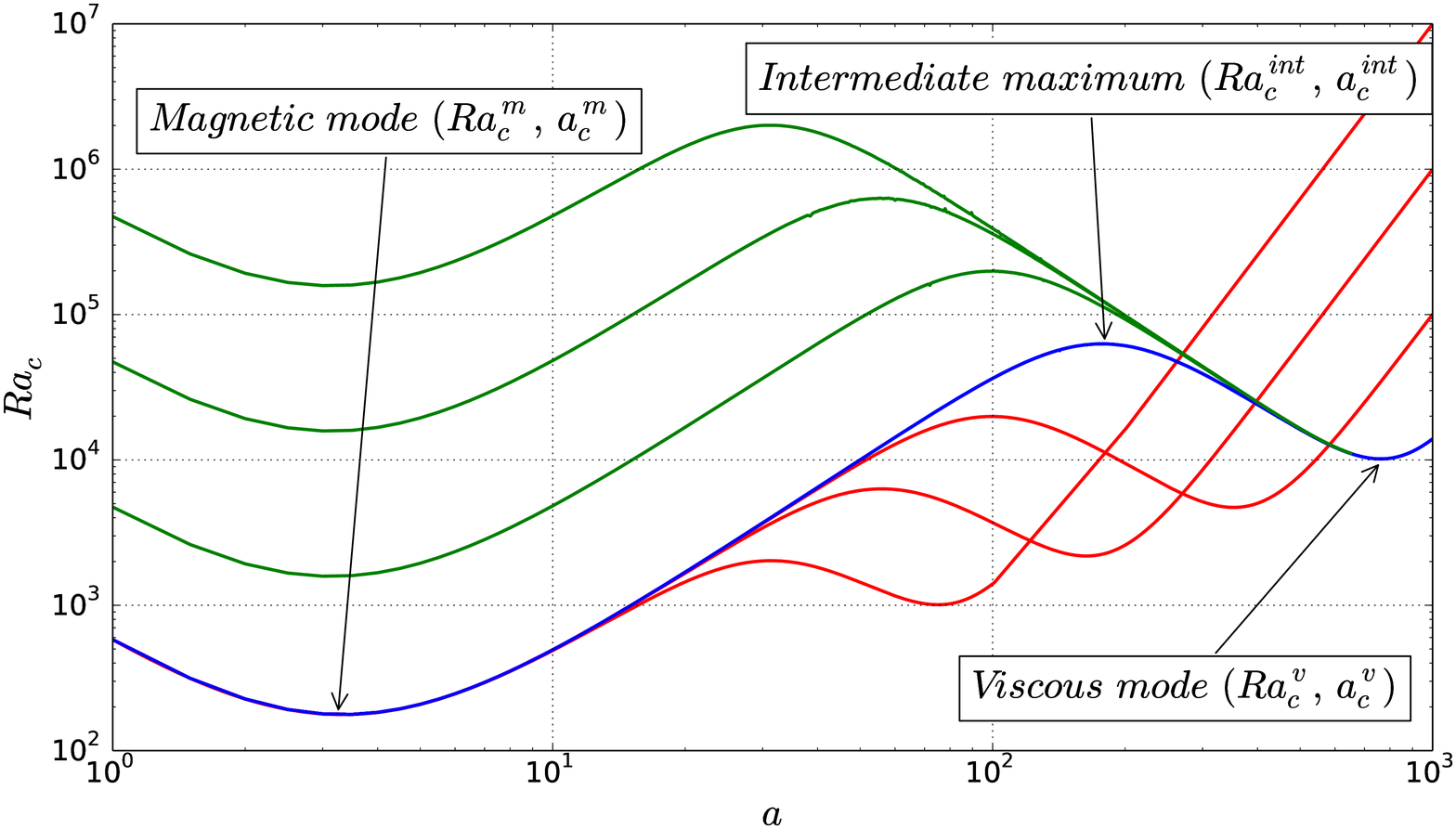}
\caption{\label{fig:Critical--with}Variation of $R\! a_{c}$ with $a$. The
blue curve's input parameters are $E=10^{-8}$ and $\Lambda=1$. The
red curve's input parameters are $E=10^{-8}$ and $\Lambda=10^{-1}\;\mathrm{to}\;10^{-3}$.
The green curve's input parameters are $E=10^{-5}\;\mathrm{to}\;10^{-7}$
at $\Lambda=1$.}
\end{figure}

In figure \ref{fig:Critical--with}, we illustrate the typical behaviour
of the critical Rayleigh number, $R\! a_{c}$ with respect to the wave number, $a$. The blue curve
corresponds to $E=10^{-8}$ and $\Lambda=1$. The green curves were
obtained for $E=10^{-8}$ and $\Lambda=10^{-1}\;\mathrm{to}\;10^{-3}$
and the red curves for $E=10^{-5}\;\mathrm{to}\;10^{-7}$ at $\Lambda=1$.
For each case, we note three specific values for $R\! a_{c}$. The
first is a minimum occurring at low $a$, its position and value depends
hardly on $E$ but is mostly controlled by $\Lambda$. As such, it
is referred to as the magnetic mode which we shall denote $(R\! a_{c}^{m},\; a_{c}^{m})$,
with $R\! a_{c}^{m}$ the magnetic critical Rayleigh number and $a_{c}^{m}$
the magnetic critical wave number. The second is a local minimum for
relatively high $a$, its position and value depending essentially on
$E$. We shall refer to it as the viscous mode $(R\! a_{c}^{v},\; a_{c}^{v})$,
with $R\! a_{c}^{v}$ the viscous critical Rayleigh number and $a_{c}^{v}$
the viscous critical wave number. Both these modes were first identified by Chandrasekhar
[\onlinecite{chandrasekhar1961hydrodynamic}]. The third feature is a local maximum located
between the two previous modes. We call this  the intermediate maximum
and denote it by $(R\! a_{c}^{int},\; a_{c}^{int})$. The corresponding  
mode is always more stable than both the magnetic and the viscous mode and 
does not reflect any mechanism driving convection. At low $E$, the value
of $R\! a_{c}^{int}$ is several orders of magnitude higher than $R\! a_{c}^{v}$
and $R\! a_{c}^{m}$.  The intermediate maximum gives a measure of how much of a 
 separation exists between magnetically controlled modes and modes controlled 
by viscosity.

\subsection{Scalings for the critical wavelength and Rayleigh number}
\label{subsec: scalings}

\begin{figure}
\subfloat[\label{fig:Scaling-for-Rac-SFM}]{\begin{raggedright}
\includegraphics[scale=0.40]{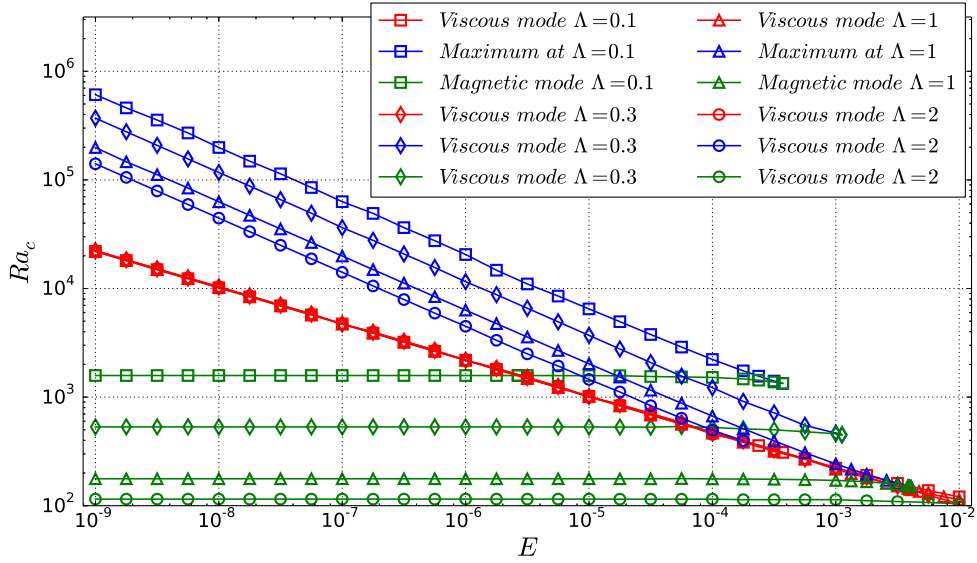}
\end{raggedright}

}

 \subfloat[\label{fig:Scaling-for-ac-SFM}]{\begin{raggedright}
\label{fig:Scaling-for-ac-SFM-1}\includegraphics[scale=0.40]{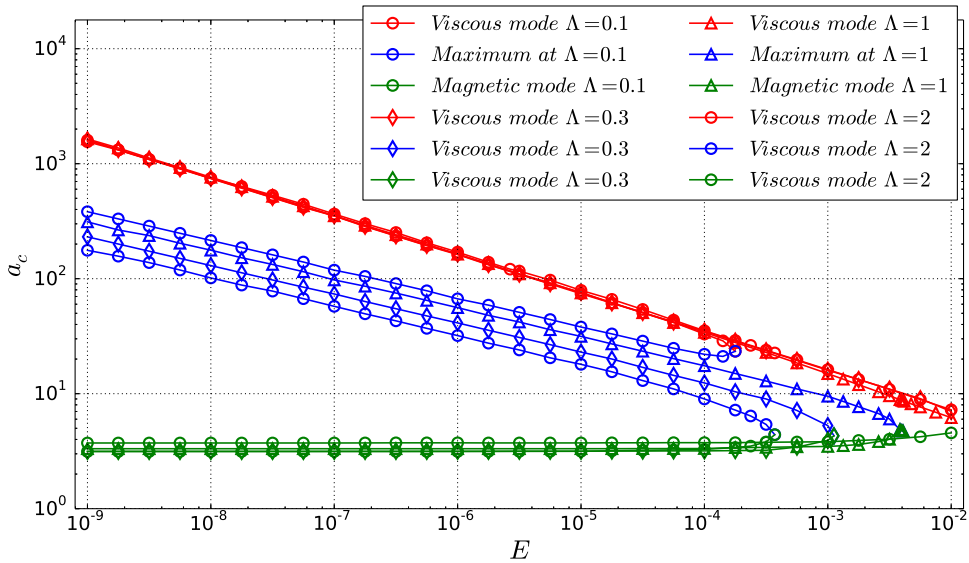}
\end{raggedright}

}

\caption{Variation of critical Rayleigh number, $R\! a_{c}$ (a) and critical wave number, $a_{c}$ (b) 
with Ekman number, $E$, for SFM boundary conditions.}
\end{figure}

In figures \ref{fig:Scaling-for-Rac-SFM} and \ref{fig:Scaling-for-ac-SFM-1},
we show the variations of $R\! a_{c}$ and $a_{c}$ with 
$E$ at $\Lambda=[0.1, 0.3, 1, 2]$ for the viscous mode, magnetic mode and for the
intermediate maximum identified in $\mathbf{A}$, with SFM boundary conditions. 
We note two important results in the limit of $E\rightarrow0$. Firstly, the scalings obtained
for the viscous modes reproduce the classical results of nonmagnetic convection, that is,
$R\! a_{c}^{v}\propto E^{-1/3}$ and $a_{c}^{v}\propto E^{-1/3}$;
and for the intermediate maximum, $R\! a_{c}^{int}\propto E^{-1/2}$
and $a_{c}^{int}\propto E^{-1/4}$. Secondly, at low $E$,
convection is initiated by the instability of the magnetic mode. On
the other hand, when $E$ increases at a fixed value of $\Lambda$,
$R\! a_{c}^{v}$ decreases while $R\! a_{c}^{m}$ remains constant,
so that a crossover value $E_{c}(\Lambda)$ exists beyond which the
viscous mode is more unstable than the critical one, and triggers
the onset of convection. Before this point is reached, the clear separation
between magnetic and viscous  progressively starts disappearing. Ultimately,
the intermediate maximum merges into the magnetic mode, at which point
both disappear, for $E=E_{D}<E_{c}(\Lambda)$.

\begin{figure}
\subfloat[\label{fig:Scaling-for-Rac-El-SFM}]{\includegraphics[scale=0.40]{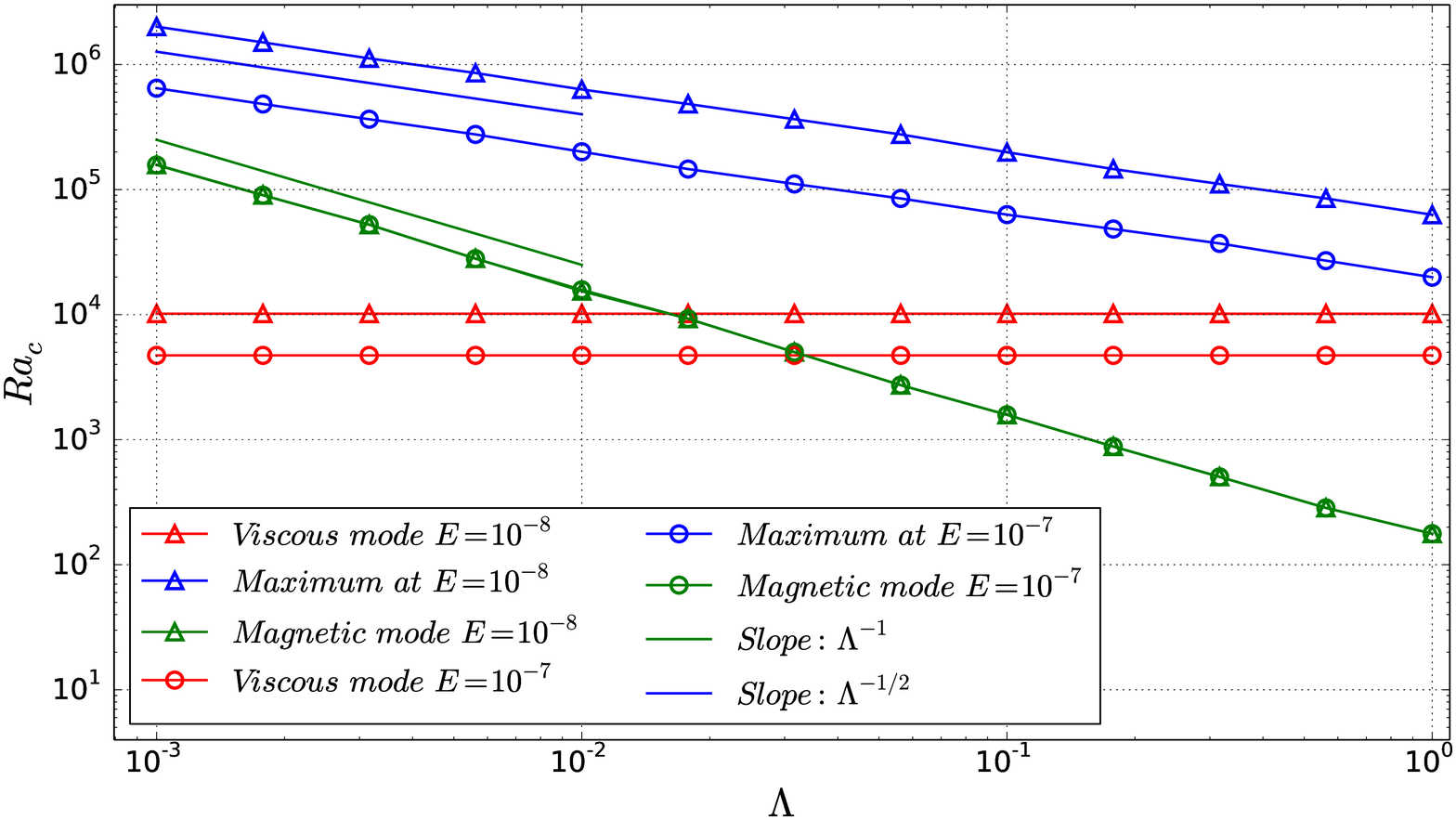}

}

\subfloat[\label{fig:Scaling-for-ac-El-SFM}]{\includegraphics[scale=0.40]{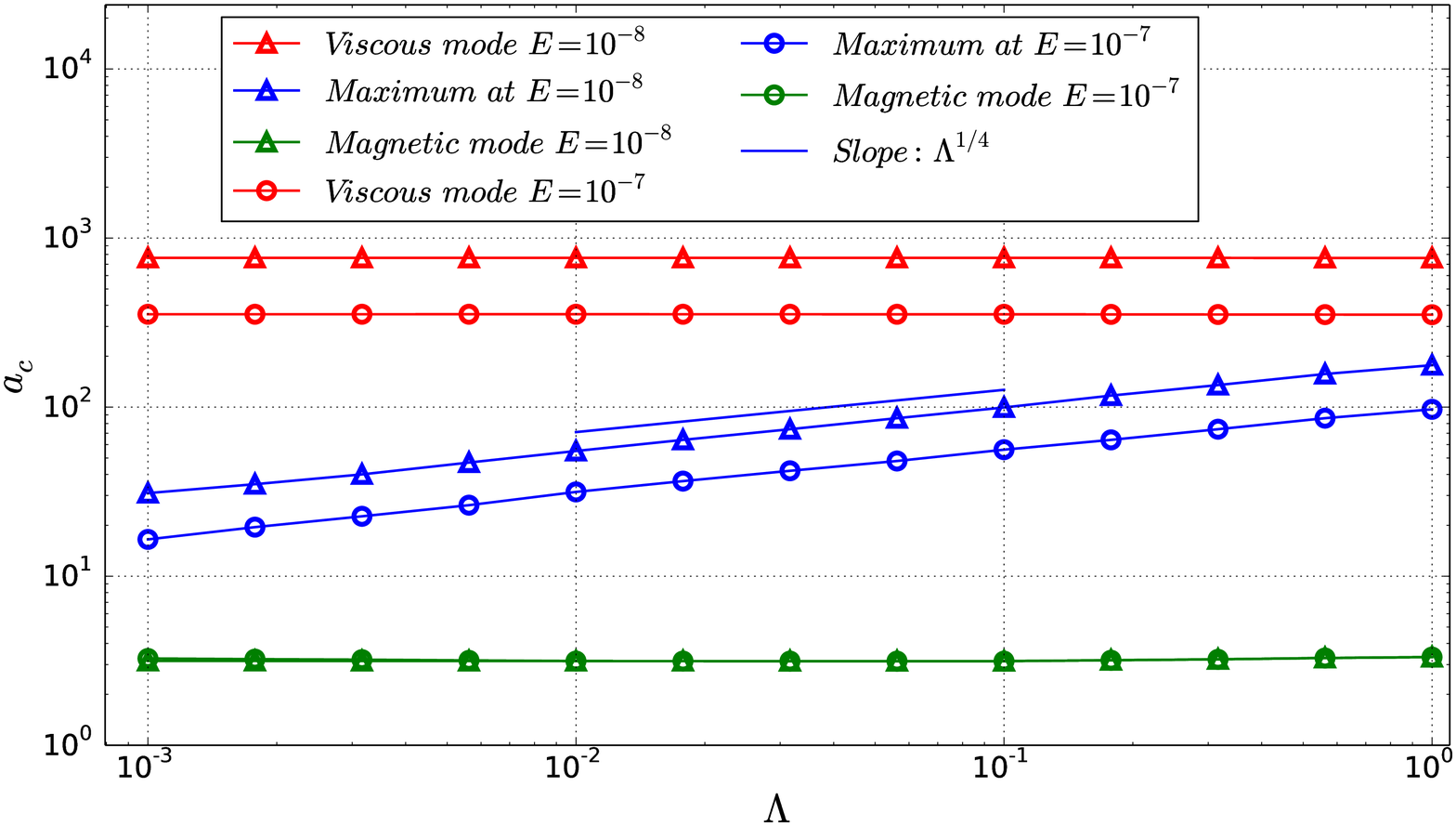}

}
\caption{Variation of critical Rayleigh number, $R\! a_{c}$ (a) and critical wave number, $a_{c}$ (b) 
with Elsasser number, $\Lambda$, for SFM boundary conditions.}
\end{figure}

In figure \ref{fig:Scaling-for-Rac-El-SFM} and \ref{fig:Scaling-for-ac-El-SFM},
we report the variations of $R\! a_{c}^{m}$, $R\! a_{c}^{int}$,
$R\! a_{c}^{v}$, $a_{c}^{m}$, $a_{c}^{int}$, and $a_{c}^{v}$ with
$\Lambda$ for $E=10^{-8}$ and $E=10^{-7}$. The Elsasser number, $\Lambda$ has been
restricted to values below $1$ which are relevant to the Earth's core.
 For higher values of the Elsasser
number, Sreenivasan and Jones [\onlinecite{sreenivasan2006azimuthal}]
showed that the Lorentz force had a stabilising effect on the
flow so that $R\! a_{c}^{m}(\Lambda)$ increases instead of decreasing
as it does for $\Lambda<1$. In the limit of $\Lambda\rightarrow0$,
we observe that the intermediate maximum scales as as $R\! a_{c}^{int}\propto\text{\ensuremath{\Lambda}}^{-1/2}$
and $a_{c}^{int}\propto\Lambda^{1/4}$. For the magnetic mode, on
the other hand, $R\! a_{c}^{m}\propto\Lambda^{-1}$ so that the separation
between magnetic and viscous modes becomes more and more pronounced
as $\Lambda$ increases. Interestingly, $a_{c}^{m}$ is practically
independent of $\Lambda$ and $E$. The crossover point at which
the magnetic mode becomes more unstable than the viscous mode can also be seen. 

\begin{figure}
\subfloat[\label{fig:Scaling-for-Rac-E-NSM}]{\includegraphics[scale=0.40]{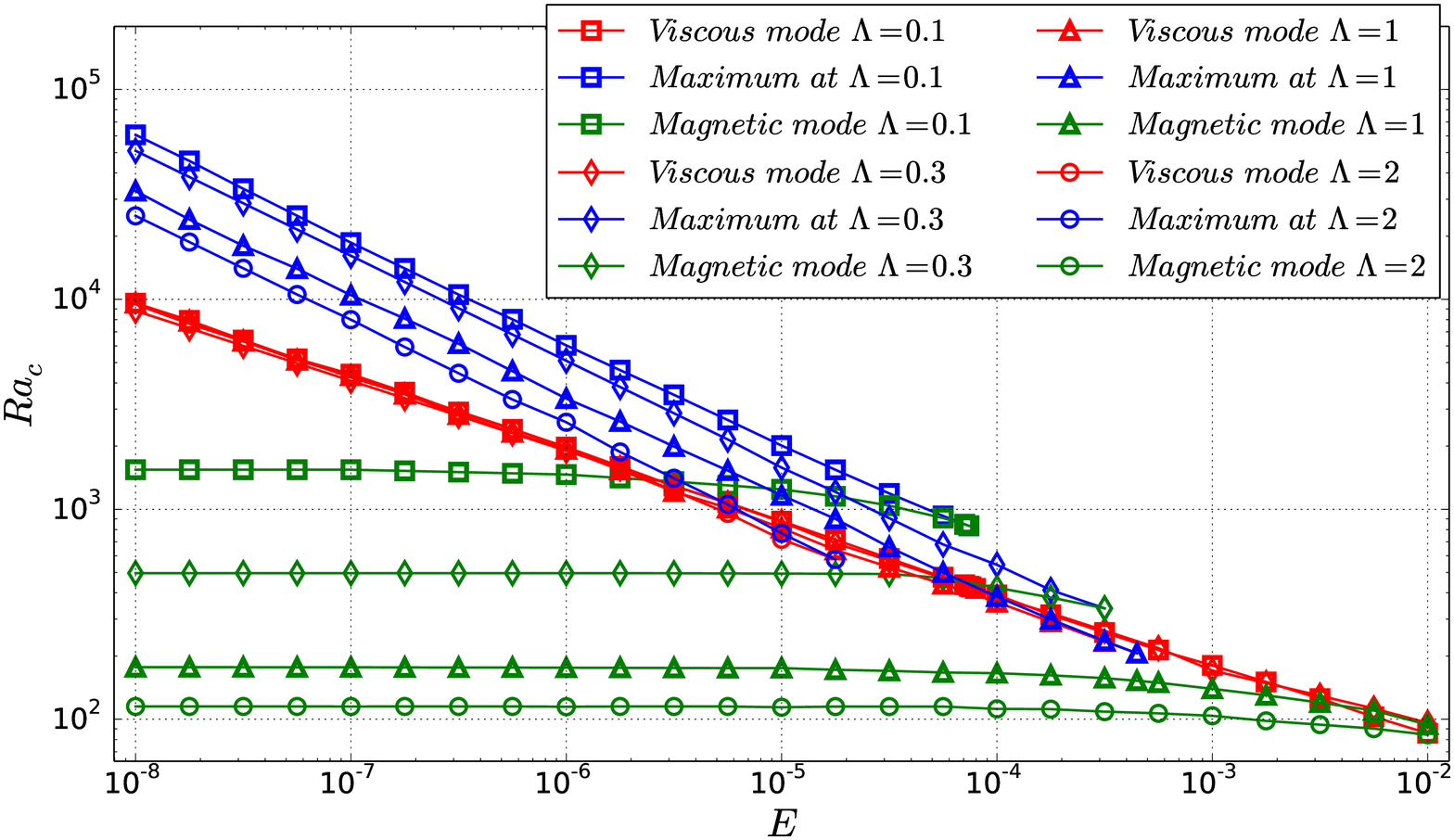}

}

\subfloat[\label{fig:Scaling-for-ac-E-NSM}]{\includegraphics[scale=0.40]{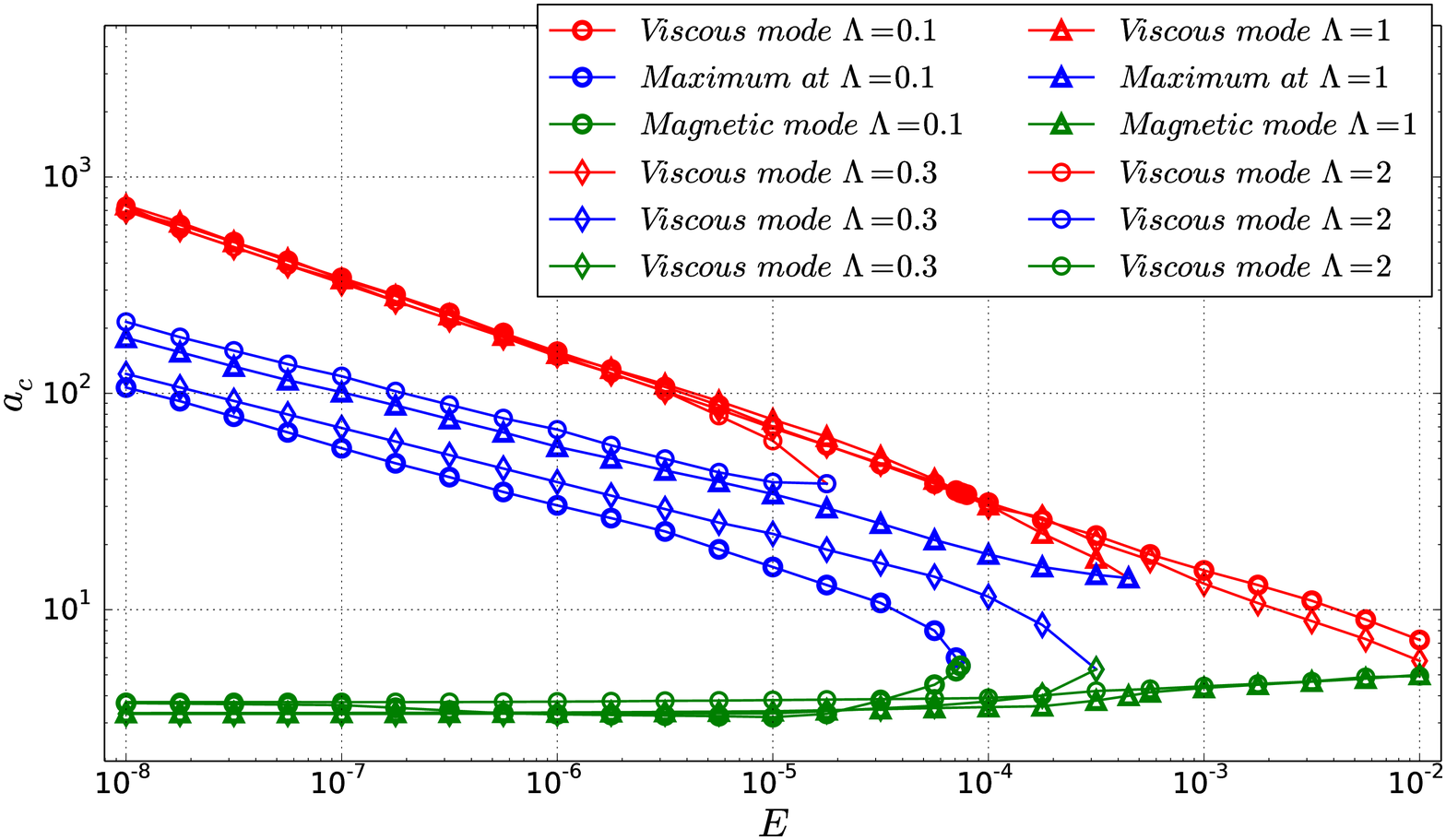}

}
\caption{Variation of critical Rayleigh number, $R\! a_{c}$ (a) and critical wave number, $a_{c}$ (b) 
with Ekman number, $E$, for NSM boundary conditions.}
\end{figure}

\begin{figure}
\subfloat[\label{fig:Scaling-for-Rac-L-NSM}]{\label{fig:Scaling-for-Rac-L-NSM-1}\includegraphics[scale=0.40]{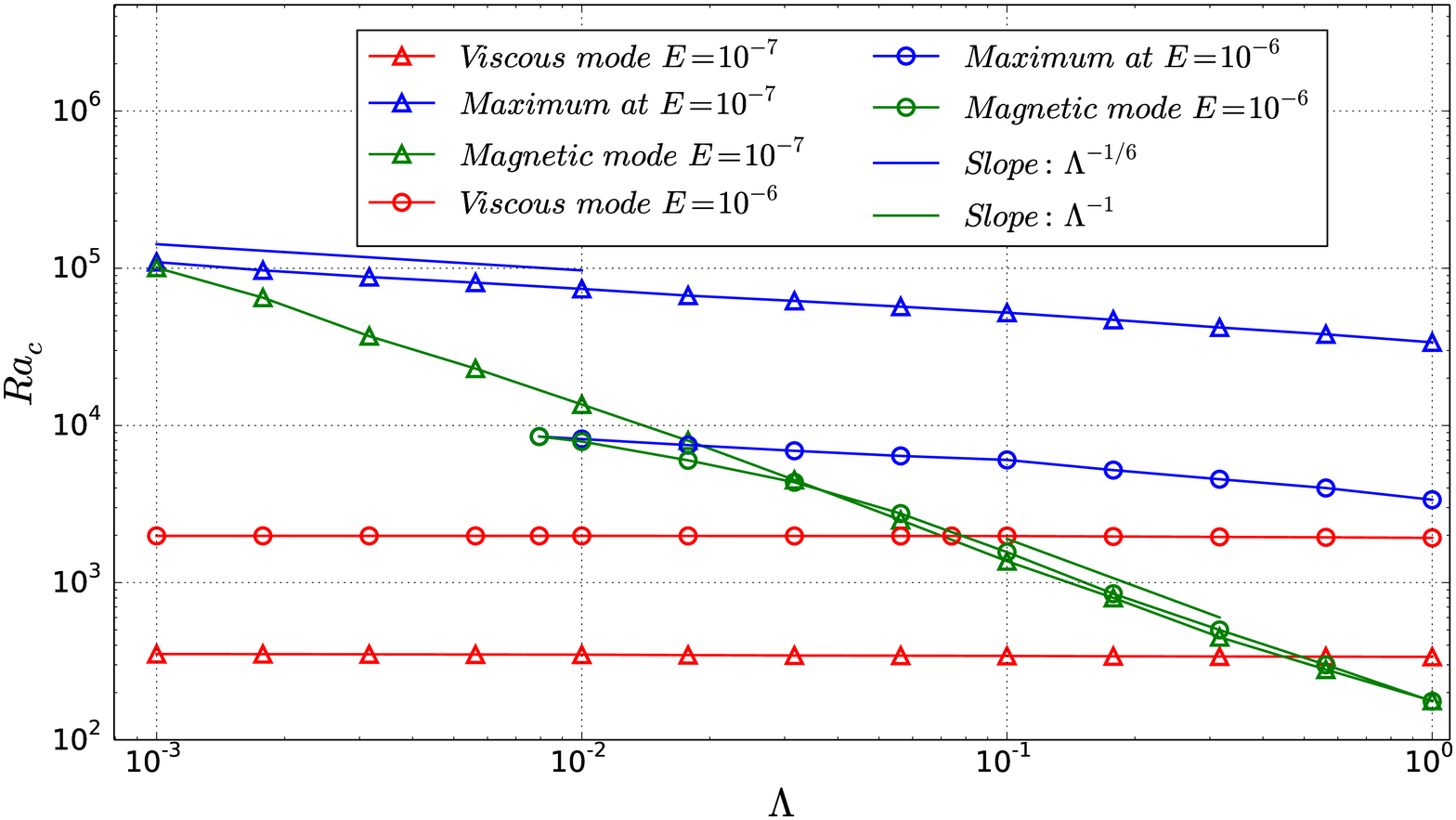}

}

\subfloat[\label{fig:Scaling-for-ac-L-NSM}]{\includegraphics[scale=0.40]{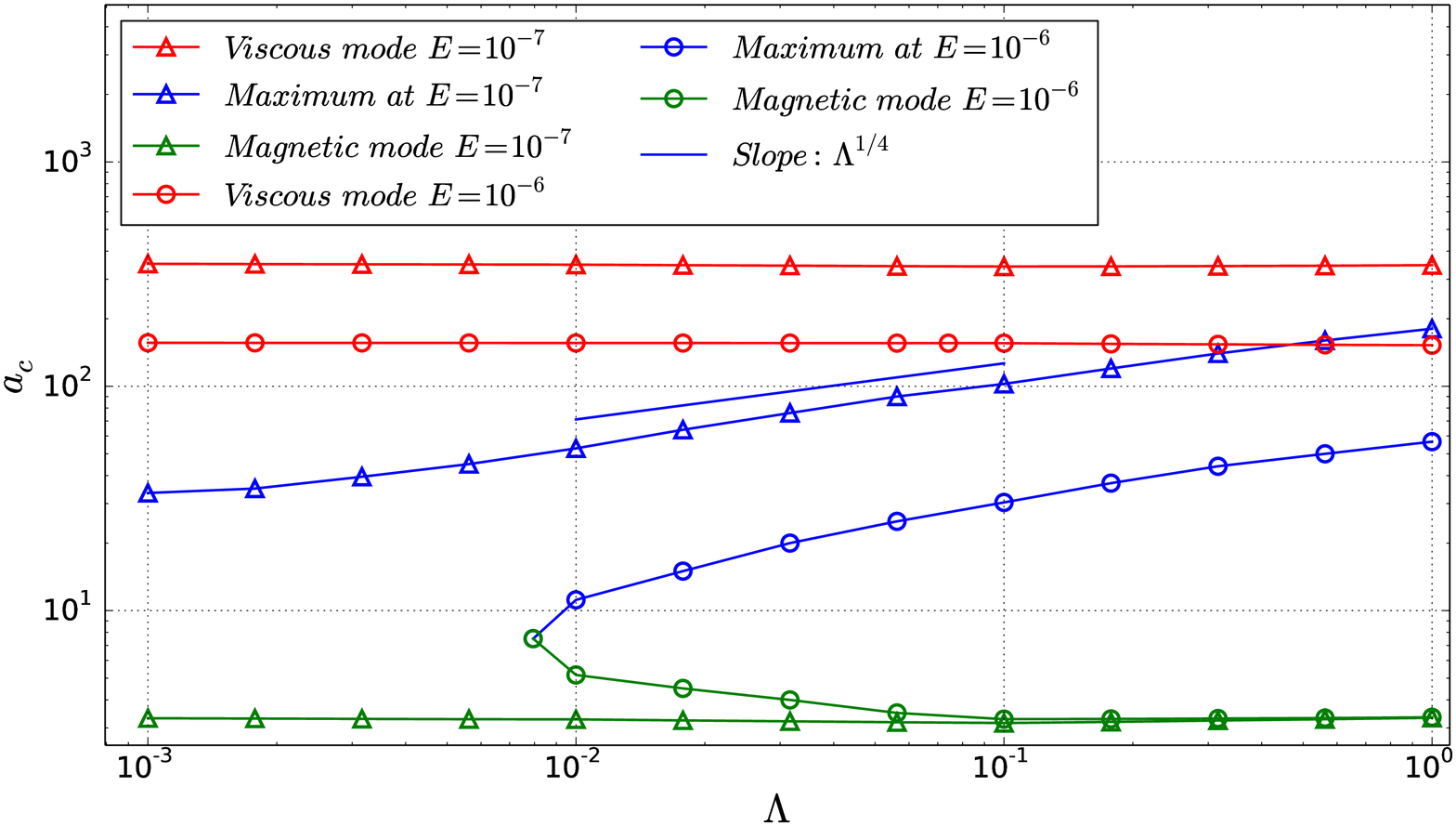}

}
\caption{Variation of critical Rayleigh number, $R\! a_{c}$ (a) and critical wave number, $a_{c}$ (b) 
with Elsasser number, $\Lambda$, for NSM boundary conditions.}
\end{figure}

%
Figures \ref{fig:Scaling-for-Rac-E-NSM}, \ref{fig:Scaling-for-ac-E-NSM}
and \ref{fig:Scaling-for-Rac-L-NSM}, \ref{fig:Scaling-for-ac-L-NSM}
present the counterparts of Figures \ref{fig:Scaling-for-Rac-SFM},
\ref{fig:Scaling-for-ac-SFM-1} and \ref{fig:Scaling-for-Rac-El-SFM},
\ref{fig:Scaling-for-ac-El-SFM} for the problem with NSM boundary
conditions. They indicate that the qualitative behaviour of the critical
Rayleigh numbers and the critical wave numbers remains the same as
in the configuration with SFM. In particular, the scalings for $R\! a_{c}$
and $a_{c}$ in the limit $E\rightarrow0$ and $\Lambda\rightarrow0$
remain valid. 

These results corroborate the findings of Sreenivasan and Jones [\onlinecite{sreenivasan2006azimuthal}]
that the boundary conditions at $z=-1/2$ and $z=+1/2$
have little influence on the onset of convection in these limits.
One important difference between the two configurations, however,
is that convergence towards the asymptotic scalings is significantly
slower with NSM boundary conditions than with SFM boundary conditions
(with a typical difference of two decades in $E$ and one decade in
$\Lambda$). With SFM boundary conditions, at high $E$ and for $\Lambda=1$, 
the intermediate maximum
merges with the viscous mode rather than with the magnetic mode. This
behaviour can be expected to take place with NSM boundary conditions
too since the wavenumbers of all three modes become closer to each
other as $\Lambda$ increases. Our results confirm the relevance
of the problem with SFM boundary conditions to the more realistic problem
with NSM boundary conditions. The small influence of the boundaries comes
as a useful feature given that simulations with NSM boundary
conditions are considerably more computationally expensive than those
with SFM boundary conditions.

\subsection{Parametric study in the $(\Lambda,E)$ space}

\begin{figure}
\begin{centering}
\includegraphics[scale=0.4]{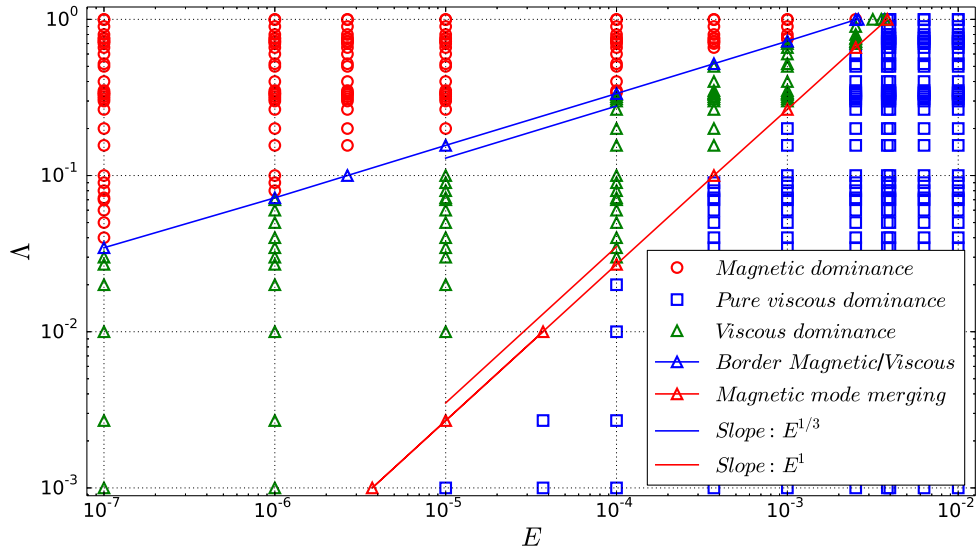}
\par\end{centering}

\caption{\label{fig:MapSFM} Characterisation of modes in the $(\Lambda,E)$ space with
Stress-free Magnetic (SFM) conditions.}

\end{figure}

\begin{figure}
\begin{centering}
\includegraphics[scale=0.4]{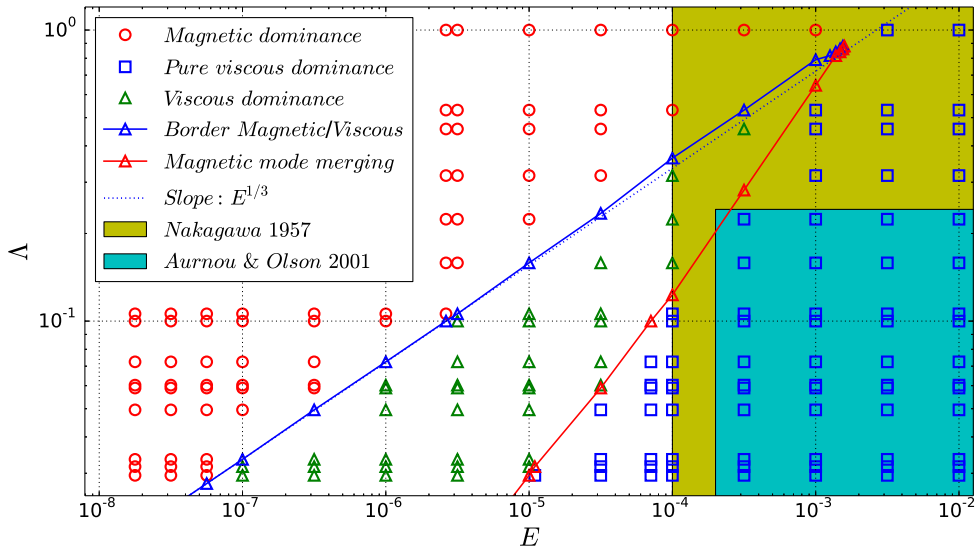}
\par\end{centering}

\caption{\label{fig:MapSFM-1} Characterisation of modes in the $(\Lambda,E)$ space with
No-Slip Magnetic (NSM) conditions.}
\end{figure}

Figures \ref{fig:MapSFM} and \ref{fig:MapSFM-1} map the mechanisms
responsible for the onset of convection in the $(\Lambda,E)$ space.
The blue squares represent the area in which only the viscous mode exists. 
The green triangles characterise the range of parameters
where magnetic and viscous modes are present but the most
unstable is the viscous one. Finally, the red circles correspond to 
regimes both magnetic and viscous modes are present but where the magnetic 
mode is more unstable than the viscous mode.
In both figures, we draw two lines, one marked with red triangles and the
other marked with blue triangles. These lines
indicate respectively the merging of the intermediate maximum with the magnetic 
mode ($R\! a_{c}^{m}=R\! a_{c}^{int}$) and the crossover line where 
$R\! a_{c}^{v}=R\! a_{c}^{int}$. In the limit of $E\rightarrow0$, these 
respectively obey the scalings:
\begin{equation}
\Lambda=270 E,
\label{eq:scal_merge}
\end{equation}
and 
\begin{equation}
\Lambda=7.22 E^{1/3}.
\label{eq:scal_crossover}
\end{equation}
%
Exponents in these laws readily follow from the scalings for 
$R\! a_{c}^{m}$, $R\! a_{c}^{v}$ and $R\! a_{c}^{int}$ obtained earlier.
These results are observed for both types of boundary conditions. In the NSM 
case, however, this asymptotic behaviour becomes only apparent for $E\sim10^{-5.5}$. In other words, the 
scaling observed at moderate $E$ with SFM boundary conditions reflects that with NSM at very 
low values of $E$. Furthermore, the fact that the shape of the curve $R\! a_{c}(a)$ is independent of the 
diffusivities $\kappa$ and $\eta$ allows us to mark out the area of parameters
investigated in experiments [\onlinecite{nakagawa1957experiments,aurnou2001experiments}]. In particular, the experiments of  [\onlinecite{aurnou2001experiments}] operates outside the viscous-magnetic transition ; which explains why these authors did not observe this phenomenon, while [\onlinecite{nakagawa1957experiments}] did. In any case, none of these experiments appear to have reached asymptotic regime of low $E$.

\section{Discussion}

In this work, we have presented a detailed parametric study of the linear stability 
problem governing the onset of plane magnetoconvection down to asymptotic 
regimes in the limit $E\rightarrow0$. This led us to the following results:

\begin{enumerate}

\item We were able to precisely verify and quantify the theoretical scalings 
for the onset of the magnetic and the viscous convection modes,
$Ra_{c}^{m}=\Lambda^{-1}$ and $Ra_{c}^{v}=E^{-1/3}$, for both NSM and SFM 
boundary conditions. 

\item Our parametric analysis led us to establish a map in the space of 
parameters $(E,\Lambda)$ and to distinguish three regions:
one where only the viscous mode exists, one where both viscous and magnetic 
modes exist but the magnetic mode is more unstable, and one where both exist 
but the viscous mode is more unstable. The crossover between 
instabilities due to the magnetic mode and instabilities due to the viscous 
one occurs for $\Lambda$=$7.22$$E^{1/3}$ in the limit $E\rightarrow0$, in agreement 
with Sreenivasan and Jones [\onlinecite{sreenivasan2006azimuthal}].

\item With NSM, this asymptotic behaviour is only recovered for $E\sim10^{-5.5}$ 
and this explains why the magnetic/viscous transition was observed in the 
experiments of Nakagawa [\onlinecite{nakagawa1957experiments}] and not in those 
of Aurnou \& Olson [\onlinecite{aurnou2001experiments}]. 

\item This asymptotic behaviour is recovered both for SFM and NSM boundary 
conditions, but attained at much lower values of $E$ for the latter than the 
former. This implies that the asymptotic behaviour found at low $E$ with NSM 
is well reproduced with SFM boundary conditions and $E$ as high as $10^{-3}$.
\end{enumerate}

Using values of  $\text{\ensuremath{\Lambda}}$ between $0.08$ and $1$,
$E=10^{-14}$ [\onlinecite{davidson2013turbulence}] and accepting the relevance 
of our simplified geometry, our results suggest that the onset of 
the convection inside the Earth's TC is magnetically controlled. 
In the same way, our analysis can be applied to Mercury,
for which $\Lambda\sim6.10^{-5}$ [\onlinecite{davidson2013turbulence}]
and $E=10^{-12}$ [\onlinecite{rudiger2006magnetic}]. Then, the asymptotic law 
(\ref{eq:scal_crossover}) suggests that the convection in Mercury's TC 
 sets off following instability of the viscous mode. 

The authors acknowledge the financial support from the Leverhulme Trust, UK (Grant RPG-2012-456), and the Royal Academy of Engineering.
\bibliography{scholar2}
\bibliographystyle{aip}

\end{document}